\documentclass[preprint]{aastex}

\usepackage{epstopdf}
\usepackage{graphics}
\usepackage{lscape}
\usepackage{natbib}
\newcommand{\zbest}{$z \sim 9.4~$}

\newcommand{\xrtposra}{\mbox{RA(J2000)=14$^{\rm h}$02$^{\rm m}$40$\fs10$}}
\newcommand{\xrtposdec}{\mbox{Dec(J2000)=$+32\degr$10$\arcmin$14\farcs6}}
\newcommand{\gemRA}{\mbox{RA=14$^{\rm h}$02$^{\rm m}$40.10$^{\rm s}$}} 
\newcommand{\gemDec}{\mbox{Dec =$+32\degr$10~$\arcmin$14\farcs20}}

\def\CIVdblt{{\rm C~}\kern 0.1em{\sc iv}~$\lambda\lambda 1548, 1550$}
\def\MgIIdblt{{\rm Mg~}\kern 0.1em{\sc ii}~$\lambda\lambda 2796, 2803$}
\def\NVdblt{{\rm N}\kern 0.1em{\sc v}~$\lambda\lambda 1238, 1242$}  
\def\OVIdblt{{\rm O}\kern 0.1em{\sc vi}~$\lambda\lambda 1031, 1037$}
\def\SiIVdblt{{\rm Si~}\kern 0.1em{\sc iv}~$\lambda\lambda1394, 1403$}
\def\AlIIIdblt{{\rm Al~}\kern 0.1em{\sc iii}~$\lambda\lambda1855,1863$}
\def\FeIIdblt{{\rm Fe~}\kern 0.1em{\sc ii}~$\lambda\lambda 2383, 2600$}

\def\AlII{\hbox{{\rm Al~}\kern 0.1em{\sc ii}}}
\def\AlIII{\hbox{{\rm Al~}\kern 0.1em{\sc iii}}}
\def\CaI{\hbox{{\rm Ca}\kern 0.1em{\sc i}}}
\def\CaII{\hbox{{\rm Ca}\kern 0.1em{\sc ii}}}

\def\CrII{\hbox{{\rm Cr~}\kern 0.1em{\sc ii}}}
\def\CII{\hbox{{\rm C~}\kern 0.1em{\sc ii}}}
\def\CIII{\hbox{{\rm C~}\kern 0.1em{\sc iii}}}
\def\CIV{\hbox{{\rm C~}\kern 0.1em{\sc iv}}}
\def\CV{\hbox{{\rm C}\kern 0.1em{\sc v}}}

\def\HI{\hbox{{\rm H~}\kern 0.1em{\sc i}}}
\def\HII{\hbox{{\rm H~}\kern 0.1em{\sc ii}}}

\def\Lya{\hbox{{\rm Ly}\kern 0.1em$\alpha$ }}
\def\Lyb{\hbox{{\rm Ly}\kern 0.1em$\beta$}}
\def\Lyg{\hbox{{\rm Ly}\kern 0.1em$\gamma$}}
\def\Lyfive{\hbox{{\rm Ly}\kern 0.1em$5$}}
\def\Lysix{\hbox{{\rm Ly}\kern 0.1em$6$}}
\def\Lyseven{\hbox{{\rm Ly}\kern 0.1em$7$}}
\def\Lyeight{\hbox{{\rm Ly}\kern 0.1em$8$}}
\def\Lynine{\hbox{{\rm Ly}\kern 0.1em$9$}}
\def\Lyten{\hbox{{\rm Ly}\kern 0.1em$10$}}
\def\HeI{\hbox{{\rm He}\kern 0.1em{\sc i}}}
\def\HeII{\hbox{{\rm He}\kern 0.1em{\sc ii}}}
\def\FeI{\hbox{{\rm Fe~}\kern 0.1em{\sc i}}}
\def\FeII{\hbox{{\rm Fe~}\kern 0.1em{\sc ii}}}
\def\FeIII{\hbox{{\rm Fe~}\kern 0.1em{\sc iii}}}
\def\MnII{\hbox{{\rm Mn}\kern 0.1em{\sc ii}}}
\def\MgI{\hbox{{\rm Mg~}\kern 0.1em{\sc i}}}
\def\MgII{\hbox{{\rm Mg~}\kern 0.1em{\sc ii}}}
\def\MgIII{\hbox{{\rm Mg~}\kern 0.1em{\sc iii}}}
\def\MgIV{\hbox{{\rm Mg~}\kern 0.1em{\sc iv}}}
\def\NaI{\hbox{{\rm Na}\kern 0.1em{\sc i}}}
\def\NV{\hbox{{\rm N}\kern 0.1em{\sc v}}}
\def\NII{\hbox{{\rm N}\kern 0.1em{\sc ii}}}
\def\NIII{\hbox{{\rm N}\kern 0.1em{\sc iii}}}
\def\NiII{\hbox{{\rm Ni~}\kern 0.1em{\sc ii}}}
\def\OVI{\hbox{{\rm O}\kern 0.1em{\sc vi}}}
\def\OI{\hbox{{\rm O}\kern 0.1em{\sc i}}}
\def\OII{\hbox{[{\rm O}\kern 0.1em{\sc ii}]}}
\def\SiII{\hbox{{\rm Si~}\kern 0.1em{\sc ii}}}
\def\SiIII{\hbox{{\rm Si~}\kern 0.1em{\sc iii}}}
\def\SiIV{\hbox{{\rm Si~}\kern 0.1em{\sc iv}}}
\def\SII{\hbox{{\rm S}\kern 0.1em{\sc ii}}}
\def\SIII{\hbox{{\rm S}\kern 0.1em{\sc iii}}}
\def\SIV{\hbox{{\rm S}\kern 0.1em{\sc iv}}}
\def\TiII{\hbox{{\rm Ti}\kern 0.1em{\sc ii}}}
\def\ZnII{\hbox{{\rm Zn~}\kern 0.1em{\sc ii}}}

\def\swift{\emph{Swift \,}}

\def\xray{X-ray}
\def\swift{\emph{Swift}}

\newcommand{\grb}{\mbox{GRB\,090429B}}
\def\simlt{\mathrel{\hbox{\rlap{\hbox{\lower4pt\hbox{$\sim$}}}\hbox{$<$}}}}
\def\simgt{\mathrel{\hbox{\rlap{\hbox{\lower4pt\hbox{$\sim$}}}\hbox{$>$}}}}

\newcommand{\gp}{\mbox{$g^{\prime}$}}
\newcommand{\rp}{\mbox{$r^{\prime}$}}
\newcommand{\ip}{\mbox{$i^{\prime}$}}
\newcommand{\zp}{\mbox{$z^{\prime}$}}

\newcommand{\nh}{\mbox{$N_{\rm H}$}} 

\newcommand{\fluence}{\mbox{$3.1\times 10^{-7}$erg cm$^{-2}$}}

\begin{document}

\title{A PHOTOMETRIC REDSHIFT OF \zbest FOR GRB 090429B}

\shorttitle{GRB~090429B at \zbest }

\author{A. Cucchiara\altaffilmark{1, 2, 3},  
        A.~J. Levan\altaffilmark{4}, 
        D.~B. Fox\altaffilmark{1}, 
        N.~R. Tanvir\altaffilmark{5},   
        T.~N.  Ukwatta\altaffilmark{6,7}, 
        E. Berger\altaffilmark{8},
        T. Kr{\"u}hler\altaffilmark{9,10},
        A.~K\"upc\"u Yolda\c{s}\altaffilmark{11, 12},
        X.~F. Wu\altaffilmark{1,13}, 
        K.~Toma\altaffilmark{1},          
        J.~Greiner\altaffilmark{9},  
        F.~Olivares~E.\altaffilmark{9}, 
        A.~Rowlinson\altaffilmark{5},         
        L.~Amati\altaffilmark{14}, 
        T.~Sakamoto\altaffilmark{7} , 
        K.~Roth\altaffilmark{15}, 
        A.~Stephens\altaffilmark{15}, 
        A.~Fritz\altaffilmark{15},
        J.P.U.~Fynbo\altaffilmark{16},
        J.~Hjorth\altaffilmark{16},
        D.~Malesani\altaffilmark{16}, 
        P.~Jakobsson\altaffilmark{17},
        K.~Wiersema\altaffilmark{5}, 
        P.~T. O'Brien\altaffilmark{5},
        A.~M.~Soderberg\altaffilmark{8}, 
        R.~J.~Foley\altaffilmark{8}, 
        A.~S.~Fruchter\altaffilmark{18},
        J.~Rhoads\altaffilmark{19},
        R.~E. Rutledge\altaffilmark{20},  
        B.~P. Schmidt\altaffilmark{21}, 
        M.~A. Dopita\altaffilmark{21}, 
        P.~Podsiadlowski\altaffilmark{22}, 
        R.~Willingale\altaffilmark{5},
        C.~Wolf\altaffilmark{22}, 
        S.~R.~Kulkarni\altaffilmark{23},
        AND P.~D'Avanzo\altaffilmark{24}                   
  	}

\email{acucchiara@lbl.gov}

\altaffiltext{1}{Department of Astronomy \& Astrophysics, 525 Davey
  Laboratory, Pennsylvania State University, University Park, PA
  16802, USA} 

\altaffiltext{2}{Lawrence Berkeley National Laboratory, M.S. 50-F, 1
  Cyclotron Road, Berkeley, CA 94720, USA} 

\altaffiltext{3}{Department of Astronomy, 601 Campbell Hall,
  University of California, Berkeley, CA 94720-3411, USA} 

\altaffiltext{4}{Department of Physics, University of Warwick,
  Coventry, CV4 7AL, UK } 

\altaffiltext{5}{Department of Physics and Astronomy, University of
  Leicester, University Road, Leicester, LE1 7RH, UK }

\altaffiltext{6}{Department of Physics, The George Washington
  University, Washington, D.C. 20052, USA} 

\altaffiltext{7}{NASA Goddard Space Flight Center, Greenbelt, MD
  20771, USA} 

\altaffiltext{8}{Harvard-Smithsonian Center for Astrophysics, 60
  Garden Street, Cambridge, MA 02138, USA}

\altaffiltext{9}{Max-Planck-Institut f\"ur extraterrestrische Physik,
  Giessenbachstr.  1, 85740 Garching, Germany}

\altaffiltext{10}{Universe Cluster, Technische Universit\"{a}t
  M\"{u}nchen,  Boltzmannstra{\ss}e 2, D-85748, Garching, Germany} 

\altaffiltext{11}{European Southern Observatory,
  Karl-Schwarzschild-Str. 2, 85748 Garching, Germany} 

\altaffiltext{12}{Institute of Astronomy, University of Cambridge, Madingley Road, CB3 0HA, Cambridge, UK}

\altaffiltext{13}{Purple Mountain Observatory, Chinese Academy of
  Sciences, Nanjing 210008, China} 

\altaffiltext{14}{INAF - IASF Bologna, via P. Gobetti 101, 40129
  Bologna, Italy}

\altaffiltext{15}{Gemini Observatory, 670 North A'ohoku Place, Hilo, HI 96720, USA}

\altaffiltext{16}{Dark Cosmology Centre, Niels Bohr Institute, Copenhagen University, 
Juliane Maries Vej 30, 2100 Copenhagen \O, Denmark}

\altaffiltext{17}{ Centre for Astrophysics and Cosmology, Science Institute, 
University of Iceland, Dunhagi 5, IS-107 Reykjav\'{i}k, Iceland}

\altaffiltext{18}{Space Telescope Science Institute, 3700 San Martin Drive, Baltimore, MD21218, USA}

\altaffiltext{19}{School of Earth \& Space Exploration, Arizona State University, Box 871404, 
Tempe, AZ 85287-1404, USA }

\altaffiltext{20}{Physics Department, McGill University, 3600 rue
  University, Montreal, QC H3A 2T8, Canada} 

\altaffiltext{21}{Research School of Astronomy \& Astrophysics, The
  Australian National University, Cotter Road, Weston Creek ACT 2611,
  Australia} 

\altaffiltext{22}{Department of Physics, Oxford University, Keble
  Road, Oxford, OX1 3RH, UK} 

\altaffiltext{23}{Department of Astronomy, California Institute of
  Technology, MC 249-17, Pasadena, CA 91125, USA} 
  
 \altaffiltext{24}{INAF-Osservatorio Astronomico di Brera, via Bianchi 46, 23807 Merate, Italy}


\begin{abstract}
Gamma-ray bursts (GRBs) serve as powerful probes of the
early Universe, with their luminous afterglows revealing the locations and
physical properties of star forming galaxies at the highest redshifts, 
and potentially locating first generation (Population III) stars.  Since GRB afterglows have
intrinsically very simple spectra, they allow robust redshifts from low
signal to noise spectroscopy, or photometry. Here we present a
photometric redshift of $z \sim 9.4$ for the {\em Swift} detected GRB\,090429B based on 
deep observations with Gemini-North, the Very Large Telescope, and 
the GRB Optical and Near-infrared Detector. 
Assuming an Small Magellanic Cloud dust law (which has been found in a majority of GRB
sight-lines), the 90\% likelihood range for the redshift is $9.06<z<9.52$, although there is
a low-probability tail to somewhat lower redshifts.
Adopting Milky Way or Large Magellanic Cloud  dust laws leads to 
very similar conclusions, while a Maiolino law does allow somewhat
lower redshift solutions, but in all cases the
most likely redshift is found to be $z>7$.
The non-detection of the host galaxy to deep limits
($Y$(AB) $\sim 28$, which would correspond roughly to 0.001$L^*$ at $z=1$) in our
late time optical and infrared observations with the {\em Hubble Space Telescope}, strongly supports the
extreme redshift origin of \grb, since we would expect to have detected any
low-$z$ galaxy, even if it were highly dusty. Finally, the energetics 
of GRB~090429B
are comparable to those of other GRBs, and suggest that its progenitor 
is not greatly different to those
of lower redshift bursts.

\end{abstract}



\section{Introduction }
\label{sec:intro}
The burst detections and rapid afterglow identifications of the
\swift\ satellite \citep{gehrels:2009lr}, combined with intensive ground-based
follow-up efforts, have confirmed some gamma-ray bursts (GRBs)  as
among the most distant objects known in the universe \citep{tfl09,sdc09},
illuminating the conditions of star formation at the earliest epochs.  As
burst detections push toward progressively higher redshifts, the mere
existence of GRBs at these times will provide 
important constraints on
models of gravitational collapse, galaxy formation,
and the early generations of stars.  
At the same time, high-quality
spectroscopy of the burst afterglows can be expected to reveal element
abundances  \citep[e.g.][]{starling05,Kawai:2006fk,bpf+06}, host galaxy kinematics, and potentially, the \HI\
fraction of the intergalactic medium (IGM), as the process of cosmic
reionization unfolds \citep[e.g.][]{Barkana:2004qy,Totani:2006bl,Tanvir:2007,McQuinn:2008uq}.

GRBs offer some advantages over other techniques for the selection and
study of distant galaxies. Most notably, they have unprecedented luminosity, 
both of the prompt emission, and afterglow \citep[e.g.][]{Racusin:2008fk,bloom09},
enabling them to provide detailed diagnostics of their environments, and 
pinpointing their host galaxies however faint. However, this utility comes at a 
price -- GRB afterglows achieve such brightness only fleetingly, and so the
time available to obtain redshifts and other information for a burst is
often very short (normally $<24$ hr). 
In order to realize the ambitions of finding bursts
at extreme redshift, and efficiently exploiting
high-redshift GRBs as probes of this early cosmic epoch, 
it is necessary to devote increasing
effort to the rapid identification of GRB near-infrared (NIR)
afterglows. In addition to workhorse NIR 
instrumentation at large observatories, a growing
number of dedicated facilities and instruments have been
commissioned, with a primary aim of rapidly locating
distant GRBs 
(e.g. PAIRITEL \citep{Bloom:2006ij}; Gamma-Ray Burst Optical and Near-Infrared Detector (GROND) \citep{Greiner:2008cq}). 
Follow-up spectroscopy of these candidates has proved several to
be at very high redshift \citep[e.g.][]{Kawai:2006fk,Greiner:2009yq}, culminating
in GRB\,090423 at $z\approx8.2$ \citep{tfl09,sdc09}.

However, in some cases rapid spectroscopy is not possible, and we must 
fall back on photometric
redshift measurements \citep[e.g.][]{Jakobsson:2006eu,haislip06}. Here again, GRBs offer some advantages over
galaxies for the application of such techniques. First, there is little intrinsic variation in 
the spectral shape of an afterglow -- it can be modeled simply as a 
power-law plus host galaxy dust extinction
Ly-$\alpha$ absorption. This is in contrast to the diverse
spectra of galaxies, which can have contributions from young/old populations (or a 
mixture), dust in complex configurations , exhibit intrinsic curvature, Balmer breaks etc, 
none of which
are a concern for GRB afterglows. Second, the identity of a GRB
afterglow is unambiguous from its fading, and thus there is no chance
of mistaking a GRB afterglow with e.g. a Galactic L or T dwarf, which
can also confuse high-$z$ galaxy searches. It has been
shown that GRB photometric redshifts are generally  
robust for these reasons \citep{Kruhler:2011rt}. Indeed, while the fundamental accuracy is
limited by the bandwidths and bands used, GRBs are
much less subject
to the ``catastrophic" failure of
photometric redshift determination, that can impact 
individual galaxy measurements.

In this paper we discuss the 
discovery and multi-wavelength follow-up of
\grb. The afterglow was 
not visible in deep early optical imaging, but
was found in deep IR observations starting $\sim 2.5$ hr after the burst. 
While spectroscopic observations were curtailed by poor weather conditions,
our photometry does
allow us to construct a spectral energy distribution (SED) for the burst, 
and to infer a photometric redshift of \zbest, making GRB~090429B
one of the most distant objects known to date.

The paper is structured as follows: in Section \ref{sec:obs} we present our
full dataset on \grb\ and the uncertainties of our photometric
measurements; in Section \ref{sec:discus} we derive our photometric
redshift, supplemented with deep host observations.  
Finally, in Section \ref{sec:sum} we
summarize our conclusions, highlighting the importance 
rapid-response NIR imaging and spectroscopic capability on large telescopes
for the study of the
early universe using GRBs. Throughout this paper we assume
$\Lambda$CDM cosmology with $H_0 = 72$ km s$^{-1}$ Mpc$^{-1}$,
$\Omega_M = 0.27$, $\Omega_{\Lambda} =0.73$, and use
a standard nomenclature to describe the variation of the afterglow flux
density as $F \propto t^{-\alpha} \nu^{-\beta}$.


\section{Observations and Analysis}
\label{sec:obs}

\subsection{{\em Swift} Observations}

The Burst Alert Telescope (BAT; \citealt{Barthelmy:2005lr}) aboard the
\swift\ satellite triggered on \grb\ at $T_0$ = 05:30:03 UT.  The
15--350\,keV light curve is composed of three distinct peaks with a
total duration $T_{90} = 5.5$\,s, and the time-integrated spectrum can
be fitted by a single power law with an exponential cut off.  The
derived total fluence in the 15--150 keV band is \fluence, with
$E_{\rm peak} = 49$\,keV. This peak energy is among the few detected
by \swift\ within the BAT bandpass.  

After 106\,s, the narrow-field instruments began their standard
burst-response observation sequence. The \xray\ Telescope (XRT;
\citealt{Burrows:2005fk}) identified an uncataloged fading source at
\xrtposra, \xrtposdec; no optical/UV counterpart was seen in the
UV-Optical Telescope (UVOT; \citealt{Roming:2005qy}) data.
The \xray\ data has been characterized using standard routines in {\sc
  Heasoft}, {\sc Xspec}, and {\sc QDP}, with the light curve fitting
process as described in \cite{ebp09}.  For some analyses, we have used
the automatic data products produced by the UK Swift Science Data
Centre \citep{ebp07,ebp09}. Our presentation of parameters derived from the
{\em Swift} data follows the convention of quoting errors at the 90\% level. The time-averaged 0.3--10\,keV
\xray\ spectrum from 97 to 29893 s after the burst is best fit by
a power-law with photon spectral index $\Gamma_{\rm X} =
2.01^{+0.16}_{-0.24}$ and with a total absorption column density of
$\nh = 10.1^{+4.6}_{-5.3} \times 10^{20}$ cm$^{-2}$, mildly
(2.7$\sigma$) in excess of the Galactic absorption of $1.2\times
10^{20}$ cm$^{-2}$; we discuss the possible significance of this
finding in Section \ref{sec:rest}.

\begin{figure*}[ht]
\begin{center}
\resizebox{14truecm}{!}{ 
\includegraphics[angle=0]{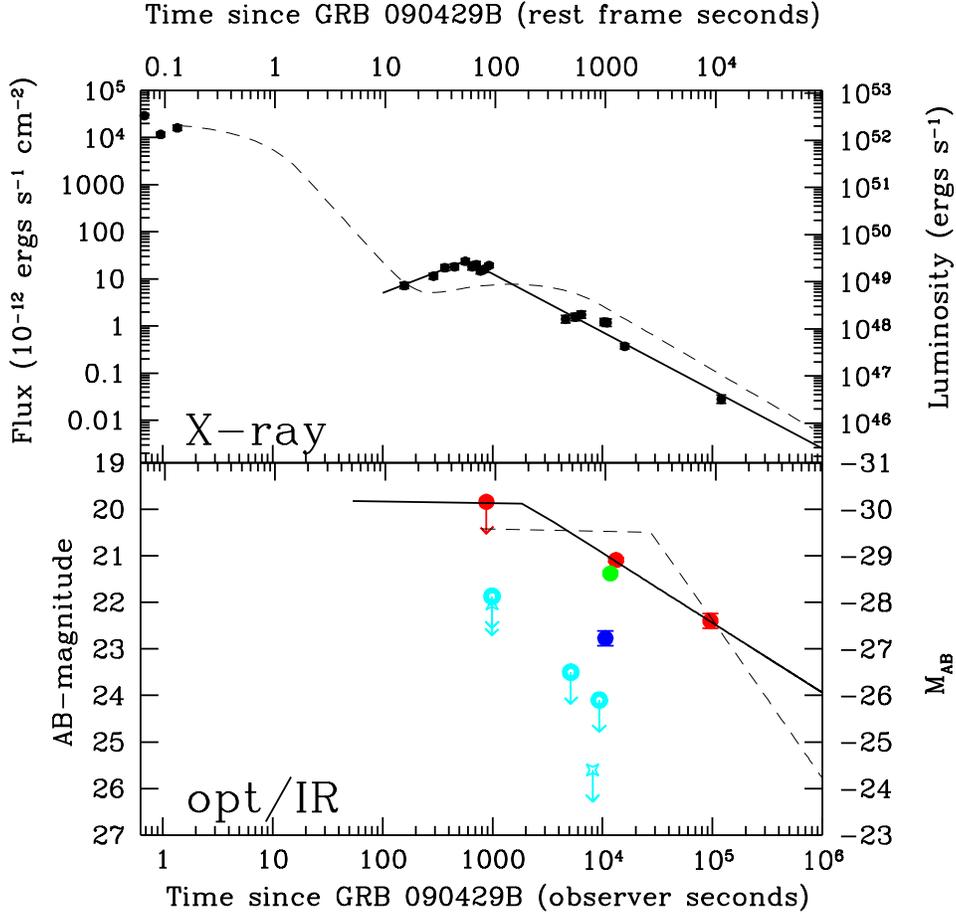} 
}
\end{center}
\caption{X-ray (top) and optical/IR (bottom) lightcurve of GRB~090429B, the left-hand and bottom axis represent the observed time and flux/magnitude, while the top and right-hand axis show rest-frame time and luminosity, respectively. The solid points in the top panel show the observed XRT data, along with a solid line representing the model.  The dashed
line represents the best fit model for GRB~090423 \citep{tfl09}
overplotted as it would appear at \zbest. The lower panel shows the optical lightcurve, along with a single power-law
fit to the (red) $K$-band points.  ($H$ and $J$ are shown as green and blue, respectively. 
For clarity we have shown only the  $i$- and $z$-band limits (cyan) in the optical). 
Additionally, the dashed line again shows the model of GRB~090423 at 
\zbest. As can be seen, the luminosity and general behavior of GRB~090429B in both X-ray and
optical is similar to that of GRB\,090423.}
\label{lc} 
\end{figure*}

The \xray\ light curve, given in Table~\ref{tab:2table2} and
illustrated in Figure~\ref{lc}, is adequately fit by a
combination of brightening and fading temporal power laws: initially,
the \xray\ flux rises with temporal index $\alpha_{\rm X1} =
-0.96^{+0.43}_{-0.52}$, referenced to the burst time; following the
peak time $T_{\rm X} = 589^{+146}_{-80}$\,s, the light curve then
breaks to a power-law decay with $\alpha_{\rm X2} =
1.20^{+0.08}_{-0.07}$.


\subsection{Optical and Near-IR Observations}
\label{sec:optical}
 
 Basic reduction steps for all optical and NIR photometry were
 performed using 
{\sc IRAF} software\footnote{IRAF is distributed by the National Optical
  Astronomy Observatory, which is operated by the Association of Universities for 
  Research in Astronomy, Inc., under
  cooperative agreement with the National Science Foundation}.
  Photometric analysis used both {\sc IRAF} and the Starlink {\sc GAIA}
software, as well as our own custom scripts.  
Errors in the sky subtraction step are estimated from multiple apertures
of size equal to that of the source aperture, placed around the field of the GRB. 

Optical images were calibrated using field stars from the Sloan Digital Sky Survey (SDSS) 
Data-Release 7 (DR7) catalog \citep{Abazajian:2009ys} in
the  and NIR images were provisionally calibrated
directly to the Two Micron All Sky Survey (2MASS) catalog, but subsequently refined as described below. 
Detections and limits on the
brightness of any associated source are presented in
Table~\ref{tab:2table1}.

 \subsubsection{ESO2.2m/GROND Observations}
 
The GROND
\citep{Greiner:2008cq} observed the field of
\grb\ simultaneously in its (dichroic and filter defined) $\gp\rp\ip\zp JHK_s$
filter set beginning 14 minutes after the \swift\ discovery
\citep{okg09}. 
No  source was detected at the \xray\ afterglow position in any of 
the seven bands:
the limits being shallower than usual due to the high airmass
for this (northern) field.
Nonetheless, the implied X-ray to optical spectral slope of
$\beta_{\rm OX} < 0.1$ implied suppression of the optical flux
relative to the X-ray, rendering GRB~090429B a ``dark" burst
under the definitions of \citet{Jakobsson:2004fr} and \citet{van-der-Horst:2009zr}.

\subsubsection{VLT Observations}

Deep $R$ and $z$-band observations were 
made with the VLT/FORS-2 camera at $\sim$60\,minutes post-burst.
Once again no optical source was visible at the position of the X-ray afterglow,
confirming that it was unusually dark, and thus a good candidate high-$z$ GRB
\citep{dlm09}.

\subsubsection{Gemini-North Observations}

Beginning roughly 2.5~hr after the burst trigger, we carried out a
series of observations from Gemini-North.  We gathered optical
\ip\zp\ imaging with the Gemini Multi-Object Spectrograph (GMOS;
\citealt{Hook:2004fj}) and NIR $JHK$ imaging with the Near-Infared
Imager (NIRI; \citealt{Hodapp:2003lr}).  GMOS observations consisted of
five exposures of 3 minutes each, per filter; NIRI observations consisted
of eight dithered positions of 60\,s each. The Gemini {\sc GMOS} and {\sc
  NIRI} packages under the {\sc IRAF} environment were used to sky-subtract,
align, and combine the images.
The NIRI images were also corrected for the small non-linearity effect seen in the
detectors\footnote{http://www.gemini.edu/sciops/instruments/niri/data-format-and-reduction/detector-linearization}.
Photometry was performed relative to
SDSS stars for the GMOS data, and relative to secondary calibrators
from GROND for the NIRI data (see Section \ref{sub:errors}).  Our photometry is
presented in Table~\ref{tab:2table1}.

\begin{figure*}[ht]
\begin{center}
\resizebox{14truecm}{!}{\includegraphics[angle=0]{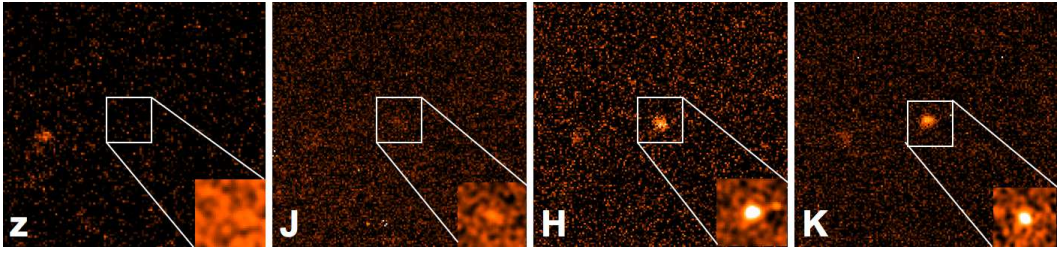}}
\end{center}
\caption{Discovery images of the GRB~090429B afterglow. The images are all obtained from Gemini-North, and show the deep non-detection in the $z$ band (which agrees with
similar observations in $griz$ obtained at GROND, $B$, $R$, $z$ obtained at the VLT, and an $i$-band image
at Gemini), coupled with the relatively bright object seen in $H$ and $K$. At \zbest, Ly-$\alpha$ lies within the
$J$ band, and explains the marginal detection at that wavelength. }
\label{disc} 
\end{figure*}

\begin{figure*}[ht]
\begin{center}
\resizebox{15.5truecm}{!}{\includegraphics[angle=0]{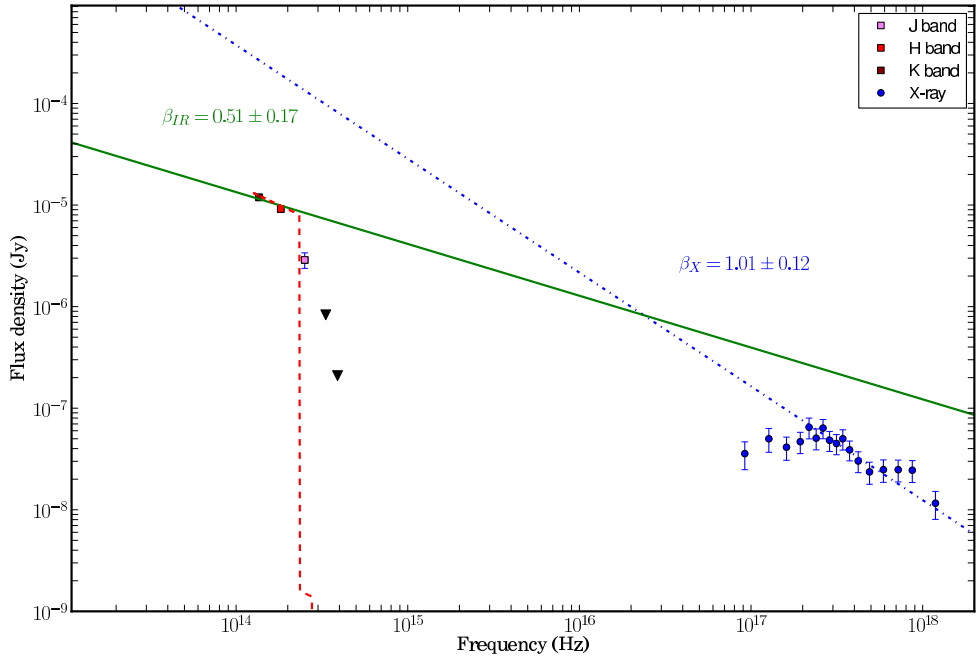}}
\end{center}
\caption{IR to X-ray spectral energy distribution at $T_0 +
  10^4$\,s can be explained by an intrinsic broken power-law spectrum. 
  The green solid line extends the IR spectral slope derived
  from the fit to the optical/NIR data, albeit that the prior on $\beta_{\rm O}$ does
  essentially fix this value.
  The blue dot-dashed line
  extrapolates the unabsorbed \xray\ spectrum to lower frequencies,
  showing that a single power-law  fails to fit the broadband SED at
  this time.  The red dashed line shows the SED for the best-fit
  extreme-redshift (\zbest) model.  \zp\ and \ip\ upper limits
  are shown as black triangles.}
\label{sed} 
\end{figure*}

While no optical counterpart was present in our \ip\ or \zp\ images,
we did identify a source within the X-ray localization in our
NIR observations. The position of the source was \gemRA,
\gemDec.  Following this discovery we attempted spectroscopic observations
from Gemini-North, however, increasing summit winds forced the closure
of the telescope and meant that
these were aborted with $<$10 minutes of useable exposure time and
no trace is visible in the observations. We obtained a second
epoch of $K$-band observations  on
April~30 UT revealed a clear fading of $\approx$1.2\,mag of the
identified source, confirming its transient nature (and corresponding to a
power-law index of $\alpha_{K} = 0.53 \pm 0.10$, shallower than the
\xray\ decay at that time).  Figure~\ref{disc} presents our Gemini imaging
data, while the lower panel of Figure~\ref{lc} shows our optical/NIR lightcurve. The
resulting SED, from X-ray to IR is shown in Figure~\ref{sed}.

\begin{figure*}[ht]
\begin{center}
\resizebox{14truecm}{!}{\includegraphics[angle=90]{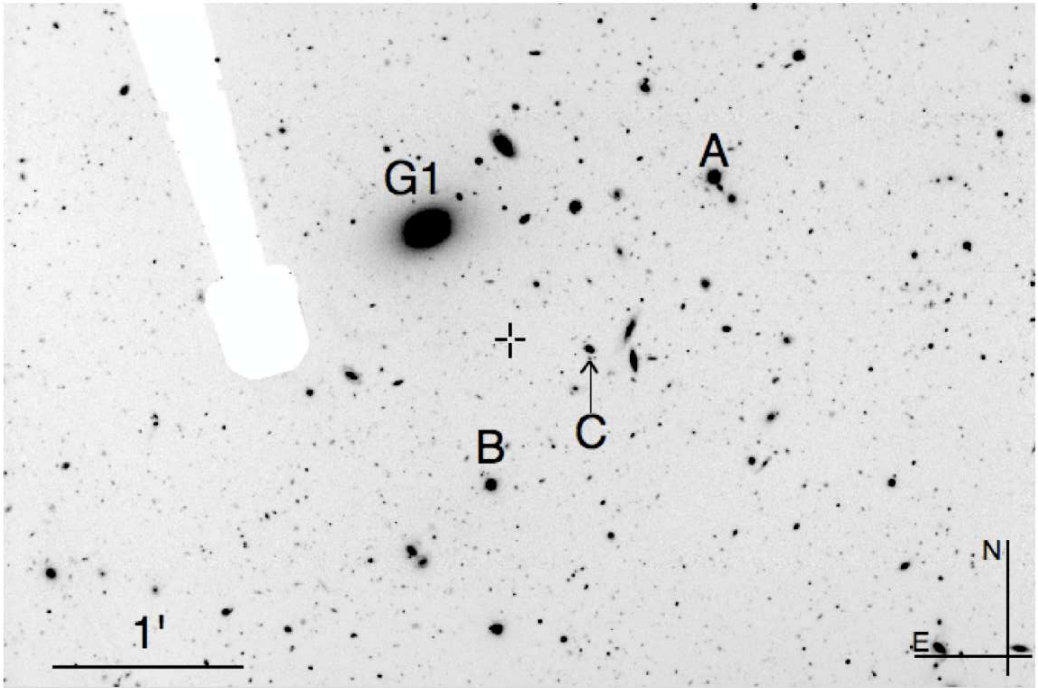}}
\end{center}
\caption{Wide-field image of the GRB~090429B field, obtained with Gemini-North 14 days after the burst. The location of the GRB is marked with a
crosshair. Additionally, we mark the positions of the three comparison stars used to refine our photometry (note that star C is faint, and lies at the end
of the marked arrow, just to the south of the galaxy), and the location of a large elliptical galaxy
(G1), which is the central galaxy of a modest cluster at $z\approx0.08$, which may provide
a modest lensing magnification. Note the silhouette of the guide probe obscures  part of the field.}
\label{wide} 
\end{figure*}

No evidence of a host galaxy is present in our images. A deep \rp-band
image of the field, taken again with GMOS under good conditions
(0.4\arcsec\ seeing) at 14~days after the GRB, is shown in Figure~\ref{wide}. 
This allows us to place a 3$\sigma$ upper limit on the host galaxy apparent
magnitude of $\rp > 27.07$\,mag. We also note in these images the presence of
a massive elliptical galaxy,  offset roughly 45\arcsec\ from the
GRB location. This galaxy has absolute magnitude $M_r\approx-21.6$
and  $M_K\approx-24.5$ \citep[$\sim L^*$; ][]{heath06}.
It appears to be the central galaxy of a modest cluster
at $z=0.079$\footnote{Redshift and $r$ magnitude of
galaxy obtained from the SDSS DR7 database; $K$ magnitude from 2MASS.},
It is likely  that this foreground structure provides some lensing boost to the
observed flux of the burst, although the relatively large impact parameter suggests
it will not be a major factor.


\subsubsection{HST Observations}

We obtained late time observations of the field of
GRB\,090429B with the {\em Hubble Space Telescope (HST)}. These were taken after the afterglow
had faded, and had the goal of finding or constraining 
the host galaxy.
We used both the Advanced Camera for Surveys (ACS) and the 
Wide Field Camera 3 (WFC3). 
Observations were obtained in {\em F606W} (broad $V-R$), {\em F105W} (broad $Y-Z$), and 
{\em F160W} ($H$): a log is given in 
Table~\ref{hstdata}. The data were reduced in the standard fashion using {\sc multidrizzle} and the {\em HST} archive ``on-the-fly" calibration.
All the images were drizzled to a common pixel scale of 0\farcs05 pixel$^{-1}$.
We ascertained the location of the burst on the {\em HST} images via relative astrometry between our first epoch $K$-band observations, and
those obtained with {\em HST}. Doing so we used a total of 11 and 10 sources in common to each frame, for ACS and WFC3 respectively. 
The resulting astrometric accuracy is 0\farcs08 ({\em F606W}), 0\farcs07 ({\em F105W}) and 0\farcs06 ({\em F160W}) respectively.  At the location of the afterglow we 
see no obvious host galaxy candidates in any of the images. 
To quantify the depths of these images we estimate the sky variance 
from a large number of background apertures ($\sim$50) placed in the field around
the target position, avoiding visible sources.
We then measure the resultant flux at the target position in an aperture of 0\farcs4 diameter,
consistent with the approaches of many groups in searching for high-$z$ galaxies \citep[e.g.][]{Bouwens:2010ly,Bouwens:2011gf}. Our fluxes are shown in Table~\ref{hstdata}. 
In addition to the measured fluxes we also show the effective AB-magnitude limits at these locations, which are equal to the measured flux density + $3\sigma$, with
an additional aperture correction to account for light missing within our small measurement apertures. These corrections are small for
ACS \citep[0.18 mag for {\em F606W}, ][]{sirianni05}, but larger for the WFC3 images (0.31 and 0.54 magnitudes for {\em F105W} and {\em F160W} respectively
\footnote{http://www.stsci.edu/hst/wfc3/phot\_zp\_lbn}).

\subsubsection{UKIRT Observations}

We obtained observations with the United Kingdom Infra Red Telescope (UKIRT), Wide-Field Camera
(WFCAM), beginning April 29 at 09:18 UT, roughly 4~hr
post-burst. Only a limited number of exposures were possible due to high wind keeping the telescope
shut much of the night.
These observations were not deep enough to reveal the afterglow, however
because the large field of view (13.6~arcmin square for each chip) includes many bright  2MASS stars 
these images allowed us to precisely determine the magnitudes of
fainter stars, which was crucial for calibrating the NIRI images (see Section \ref{sub:errors}).
Pipeline reductions were performed by the Cambridge Astronomical Survey Unit 
(CASU\footnote{http://casu.ast.cam.ac.uk/}).

\subsection{Precise Photometric Calibration and Uncertainties}
\label{sub:errors}

Since our photometric redshift analysis will depend critically on the
accuracy of our photometry, we took particular care in both  the calibration
and estimates of photometric uncertainties.

The Gemini-North/NIRI detections are crucial, but also difficult to analyze since the field of view is small
(2~arcmin on a side) and there is only one 2MASS star (namely star B in Table~\ref{crosscalib},
and Figure~\ref{wide})
that is in all the sub-exposures of the nine-point dither pattern.  
There is another 2MASS star (A in Table~\ref{crosscalib}) which appears on 
two of the sub-exposures, and we used this as a double check on the derived photometry.
Both these stars are toward the faint end of the 2MASS catalog and  have relatively large 
photometric uncertainties. To overcome this we used the wide-field UKIRT/WFCAM
and ESO2.2m/GROND $JHK$ images (both of which were obtained close in time to the Gemini observations), which were very precisely
calibrated using many bright 2MASS stars, to obtain more accurate magnitudes for these reference stars.
The two independent determinations were consistent with each other within their respective 
calculated errors (typically 0.01--0.02 mag), and we therefore formed a weighted average to obtain
our best estimates of the  NIR magnitudes, as shown in Table~\ref{tab:2table1}.

Magnitudes for the afterglow were measured relative to star B, although this procedure 
was further complicated by the fact that
the point-spread-function (PSF) was found to change across the frame resulting in  the core of the reference star becoming noticeably
extended when it was close to the southern edge of the detector, as it was in some sub-exposures.  This precluded small
aperture (5 pixel radius, $\approx0\farcs6$ ) photometry for these exposures, so in such cases we used a fainter star 
(C in Table~\ref{tab:2table1}) closer to the GRB position as a secondary
reference, having determined its magnitude relative to  star B using those frames where it was not near the edge.
We note that a small aperture was required to maximize the signal to noise for the afterglow, and
that profile-fitting photometry was deemed inappropriate due to the small number of bright stars
available to define the PSF.

The magnitudes (and errors) for the afterglow in each band were then determined from an error weighted mean of the different sub-exposures.
Finally, we converted to flux density using a recent NIR spectrum of Vega \citep[see ][]{Bohlin:2007ul} which resulted in values that are
2\%--3\% higher for our passbands
than found using the conversion in Table 7 of \citet{hewett06}.
These are the flux densities reported in Table~\ref{tab:2table1}, although we note that when we come to the photometric redshift analysis (below)
we fit in counts rather than flux, to allow for the different spectral shapes of the afterglow and comparison star.

Since the optical observations provided only upper limits 
the overall fit is not strongly sensitive to the precision of the optical photometry. 
However, in this case, the field of GRB~090429B lies fortuitously within the SDSS survey area, 
and our most constraining optical
limits (from GMOS) are obtained in the same filter set. This allows a precise photometric calibration of these images. 
For our Very Large Telescope (VLT) observations we calibrate the field using SDSS observations
and the transforms of \citet{jester}.  These latter values were confirmed as reasonable using archival zeropoints.


\section{Results and Discussion}
\label{sec:discus}


\subsection{Temporal Behavior}

Since our observations were taken over several hours, temporal variations
in the afterglow luminosity could effect our analysis.  Unfortunately we have
rather little handle on the variability at optical/IR wavelengths.
Although most GRBs begin power-law decline in luminosity fairly early,
in some cases flat or even increasing luminosity can be seen for a period
of time \citep[e.g.][]{rykoff09}.
A rapidly fading afterglow (similar to those
commonly observed) would imply even more stringent upper limits in our
(earlier) blue band filters since the extrapolation to a common time would
yield a more extreme limit on
color index in e.g. $z-H$. Alternatively, the more unusual case
of a rapidly rising afterglow would yield somewhat weaker constraints
since the non-detections in the bluer bands could be ascribed to the brightening
of the afterglow in the time frame between the optical and IR observations. 
However, this is countered by the fact that such a rising afterglow would also imply an
even bluer $H-K$ color, more difficult to attain with extinction.  
In this scenario, the rising of the afterglow
may offer some support for a high-redshift scenario, since the time dilation 
at $z \sim 9$ would result in a forward shock which takes a factor of $\sim 10$ 
longer to reach maximum than at $z \sim 0$. 

In an attempt to constrain the temporal slope of the optical afterglow
we first perform photometry on the individual NIRI frames. We find no statistically significant
variation over the $\sim 10$ minute time frame of these observations, implying
that the afterglow is not varying especially rapidly. Second, we 
utilize acquisition images taken prior to the
aborted NIRI spectroscopic observations. These suggest a minor brightening
of the afterglow between 13000\,s and 17000\,s after the burst 
($0.3 \pm 0.2$ mag, corresponding to $\alpha\approx-1.0\pm0.7$). However, these observations were 
obtained at a single dither position, and contained substantial persistence.
Hence, we can accurately remove neither sky nor dark current and the resulting
observations contain large variations in the sky on relatively short
length scales. Thus, we caution against their use for detailed photometric
work, aside from noting that suggest that the afterglow is neither rising, 
nor falling at an unusually rapid rate. 
We gain a much better handle on the decay between the first and second
night observations, which gives $\alpha\approx0.6\pm0.1$, but of course
this may not apply during the first few hours post-burst.

On balance, then, we favor photometric fitting
in which  the observed magnitudes are assumed to be constant over the
period of the early observations (i.e. we
assume $\alpha = 0$). This is consistent with the relatively flat X-ray behavior
between 1 and 3 hr post-burst (Figure~\ref{lc}). For completeness, 
we have included a single power-law temporal decay as a possible
parameter within these fits, and confirm that for any reasonable 
slope $-1 < \alpha < 1$ our results are broadly insensitive to the
assumed value of $\alpha$ (see below).
To avoid extrapolating over too wide 
a range of times, and to counter against unusual afterglow
behavior (which is normally most notable in the first few hundred seconds in the rest frame) we also 
fit only data taken after 4000\,s. The inclusion of earlier
data would further strengthen our results if we assumed  the afterglow to be
decaying, but would make minimal difference to the fit if we assumed a flat
or rising light curve (since the early limits are shallower than those obtained at
later times).

\subsection{Photometric Redshift Analysis}
\label{sub:photoz}

Here we attempt to derive the redshift of GRB~090429B via our broadband photometry of its afterglow.
The absence
of any detections within the optical window, if interpreted as the signature of high
redshift, immediately implies $z>6.3$. Similarly, if we interpret the red $J-H$ color
of the afterglow as indicative of a Lyman-break lying within the $J$-band, the inferred
redshift is $8.0 < z < 10.5$.

To obtain stronger constraints on the redshift of GRB\,090429B 
we performed the following analysis.  We considered just the seven deepest
observations, namely those obtained at the VLT and Gemini-North on the first night with
filters redder than 6000\,\AA. We assume initially
there is no temporal variation over the course of observations, although including 
plausible variability
within our fits also confirms that our results are broadly insensitive to this assumption (see below).
After correcting for Galactic foreground extinction \citep[$E_{B-V}=0.015$; ][]{sfd98} we fitted these 
flux density measurements
with a grid of simple models for the SED of the afterglow.  
The errors are likely  to be Gaussian distributed, to good
approximation, since the uncertainties are dominated by background subtraction
(although we also included zero-point calibration uncertainties in the modelling),
and therefore used minimum-$\chi^2$ fitting.

\begin{figure*}[ht]
\begin{center}
\resizebox{14truecm}{!}{\includegraphics[angle=270]{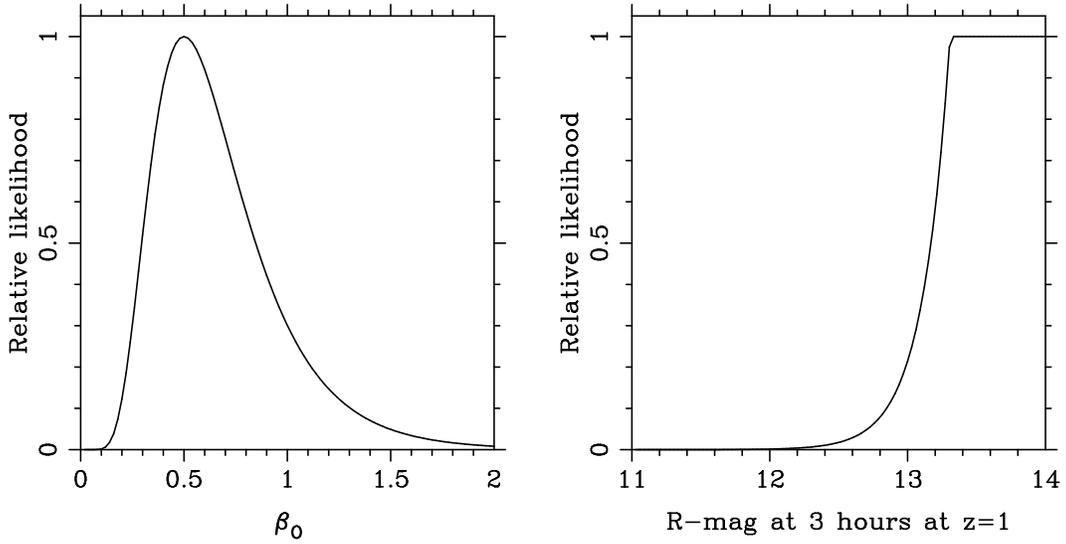}}
\end{center}
\caption{Input priors adopted for our photometric redshift fitting. 
[Left panel:]
In the relativistic fireball  model, the intrinsic spectral slope in the optical 
should lie between $\beta_{\rm X}$ and 
$\beta_{\rm X} - 0.5$ (plus the associated measurement errors).
To achieve this
we use a lognormal distribution centered at 0.5 (since
there does appear to be a break between the optical and X-ray, see Figure~\ref{sed}). This is a relatively weak prior and simply avoids extreme values of $\beta$. 
Right panel:
The second prior is on the intrinsic optical afterglow luminosity, 
and impacts solutions that would result in an unreasonably bright luminosity (it is not bounded at the  faint end, and hence the low-redshift solutions are unaffected). 
It is therefore based on the empirically observed upper envelope of afterglow luminosities.
The primary impact of this
prior is to disfavor moderate ($A_V >3$) scenarios at high redshift $(z>7)$, where the burst would have been 
more luminous than any other known afterglow. }
\label{priors} 
\end{figure*}

Specifically, the 
model was a simple power law, with the spectral index, $\beta_{\rm O}$, and overall normalization as free parameters. 
The grid of models spanned a range in redshift of $0<z<12$ and rest-frame $V$-band extinction of $0<A_V<12$.
Beyond $z \sim 7$ we are effectively fitting only three data points so there exists a degeneracy between
extinction ($A_V$) and  $\beta_{\rm O}$.  
We therefore include a weak prior for the probability distribution of the
value of $\beta_{\rm O}$, (see Figure~\ref{priors}) which is modeled
as a log normal  with a peak likelihood at  $\beta_{\rm O} = 0.5$ and a width such
that the relative likelihood is 50\% of the maximum from about $0.3<\beta_{\rm O}<0.85$.
This is
physically motivated since it allows values of $\beta_{\rm O}$ over a broad range,
comparable to the range usually observed \citep[e.g.,][]{schulze10}, but in particular
prefers $\beta_{\rm O} = \beta_{\rm X} - 0.5\approx0.5$, as would be expected if there
was a cooling break between the X-ray and optical regimes \citep[e.g.,][]{sari98}. 
The plausibility of such a cooling break is clear from Figure~\ref{sed}, while
relaxing this assumption does not lead to any viable fits at low redshift. Added to this
was absorption due to neutral hydrogen in the IGM \citep{mad95},
(neutral hydrogen in the host was taken to have a typical column density of $10^{21}$\,cm$^{-2}$, although the results are
insensitive to the exact number assumed), and extinction due to dust.  We experimented with
several dust laws, from the Milky Way (MW), Large Magellanic Cloud (LMC) and
Small Magellanic Cloud (SMC) \citep{pei92}, as well as the extinction law of 
\citet{maiolino04}.

We also impose a weak prior on the intrinsic luminosity of the optical afterglow
(Figure~\ref{priors}).
Studies such as that of \citet{kann2010}, indicate that there is an upper
envelope to the (broad) distribution of GRB optical afterglow luminosities.
We therefore apply a prior which is flat below this envelope and cuts
off exponentially at brighter luminosities, although the cut-off is slow enough to allow
a reasonable probability that the luminosity could be somewhat higher.

\begin{figure*}[ht]
\begin{center}
\resizebox{14truecm}{!}{\includegraphics[angle=270]{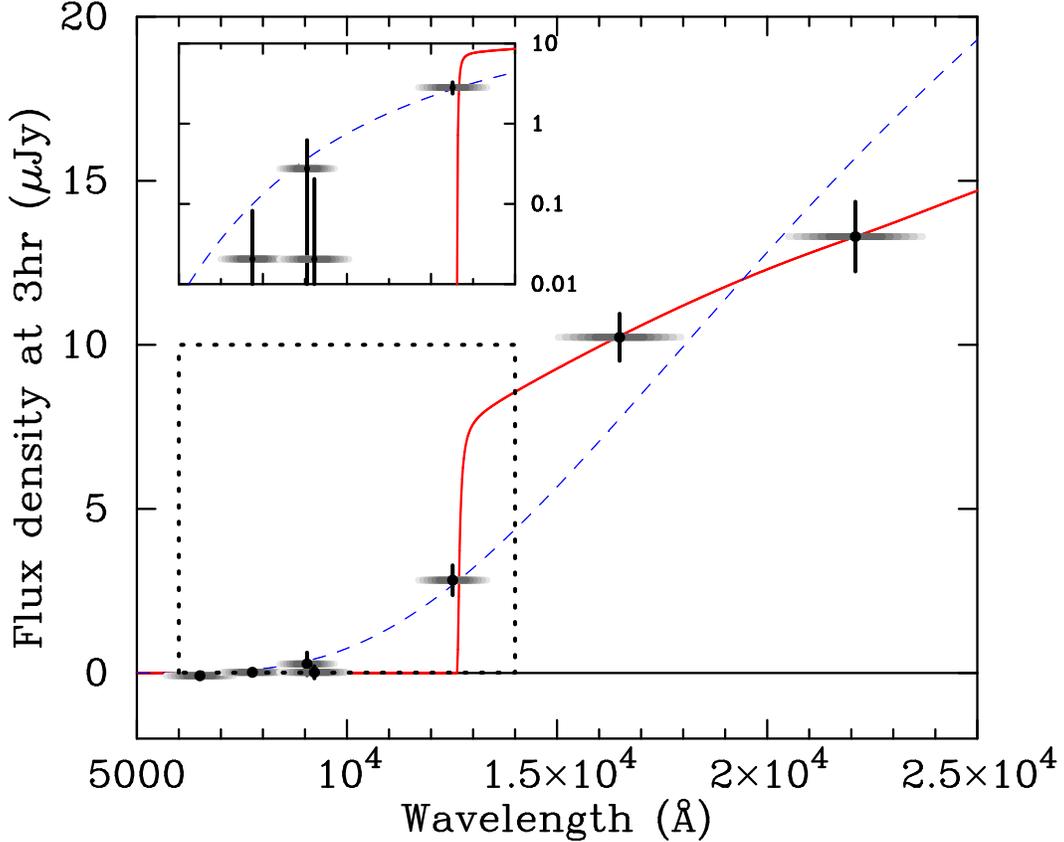}}
\end{center}
\caption{Spectral energy distribution of the GRB~090429B afterglow formed by 
extrapolating our observed photometry to 3 hr post-burst assuming the magnitude
remains constant, i.e. $\alpha=0$ (for varying $\alpha$ fits see Figure~\ref{contpanel}).
The vertical error bars represent 1$\sigma$ uncertainty, and the horizontal shaded
bars illustrate the widths of the broadband filters.
The best fit model ($\chi^2/dof = 1.76/3$) to the data points
is shown as the solid red line, the parameters being redshift $z=9.36$,
rest-frame extinction $A_V=0.10$ and intrinsic power-law slope $\beta_{\rm O}=0.51$.
The inset simply replots the short wavelength part of the figure (indicated by a dotted
box) on a logarithmic flux density scale, to more clearly show the constraints
from the optical measurements.
An alternative low-redshift ($z\approx0$), high extinction ($A_V=10.6$) model
is shown as a dashed blue line, but in fact is
formally ruled out at high significance ($\chi^2/dof = 26.2/4$). 
}
\label{sedfit} 
\end{figure*}

\begin{figure*}[ht]
\begin{center}
\resizebox{14truecm}{!}{\includegraphics[angle=270]{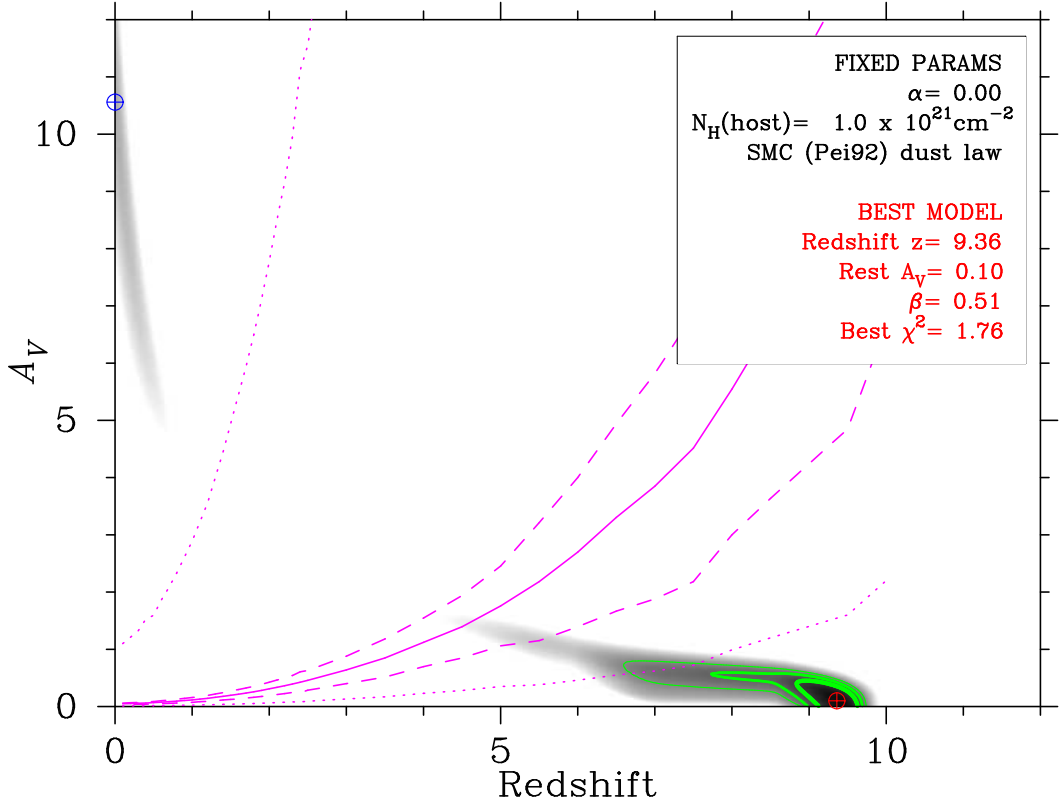}}
\end{center}
\caption{Confidence contours on a parameter space of
redshift and host galaxy extinction 
for the GRB~090429B afterglow, 
for our favored set of prior assumptions
(green contours are 90\%, 99\% and 99.9\% confidence). 
The gray scale shows likelihood down to much lower levels, formally $\sim10^{-7}$.
All fits at $z<7.7$ are ruled out at $>99$\% confidence, and while fits can be 
found at $z \sim 0$ they are markedly worse than the high-$z$ solutions. The best $z<5$ solution (formally at $z=0$) is marked with the blue cross and requires $A_V \sim 10$, 
and is also disfavored by the lack of any host galaxy to deep limits, and the inconsistency of the required $A_V$ with the hydrogen column density
measured from the X-ray afterglow. To illustrate this the best-fit $N_{\rm H}$ from the X-ray spectrum
is converted into $A_V$ and plotted onto the contour plot as the purple lines 
\citep[dashed lines show the 90\% error range, and the dotted lines show the limits of the systematic 
error due spanning the range of gas-to-dust ratios reported by][]{schady10}.
As can be seen the $A_V$ inferred from the X-ray, and that required from 
the photometric redshift fit are inconsistent at low redshift, but broadly consistent with the
high-$z$ fit. }
\label{cont} 
\end{figure*}

In fact, rather than fitting directly to the flux densities, we integrated our model spectra over response
functions and compared the counts obtained by integrating an approximate spectrum
of the comparison star.  The response functions were obtained from the measured filter
transmission curves, multiplied by a typical atmospheric absorption curve generated
by ATRAN\footnote{http://atran.sofia.usra.edu/cgi-bin/atran/atran.cgi}.
Going to these lengths effectively corrects for the small
difference in the SED shape of the afterglow from the reference star, although again 
the conclusions are not greatly affected.
In Figure~\ref{sedfit} we show the photometric data points and
the best-fit model for the afterglow spectrum assuming that the afterglow did not
evolve temporally during the first 3 hr (see below) and that the dust is similar to
that in the SMC, which has frequently been found to be
a good approximation to the dust laws along many other GRB sight lines \citep[e.g.,][]{schady07,schady10}.
We also show the best fit low-$z$ model (as it happens $z\approx0$), 
which is formally ruled out at high confidence.
In Figure~\ref{cont} we plot contours of $\chi^2$ over a grid of models
spanning a range of redshift and rest-frame $V$-band extinction, $A_V$.  
The red-cross shows the best fitting model, which has $z=9.36$ and extinction $A_V=0.1$,
although the 99\% confidence contour runs as low as $z\approx7.7$ if there is a modest
amount of dust (rest-$A_V\sim0.5$) in the host.
Marginalizing the likelihood (which we define  $\mathcal{L}\propto{\rm exp}(-\chi^2/2)$)
over $A_V$ (assuming a flat prior) indicates
a 90\% likelihood range of $9.02<z<9.50$.
There is no solution at lower redshifts ($z<7$) which is not ruled out at  $\gg$99.9\% level:
and the best fit at low redshift ($z\approx0$ as it happens, as shown by the blue cross) requires very high extinction
of $A_V\approx10$.

\begin{figure*}[ht]
\begin{center}
\hspace{-0.2cm}
\resizebox{16.5truecm}{!}{\includegraphics[angle=90]{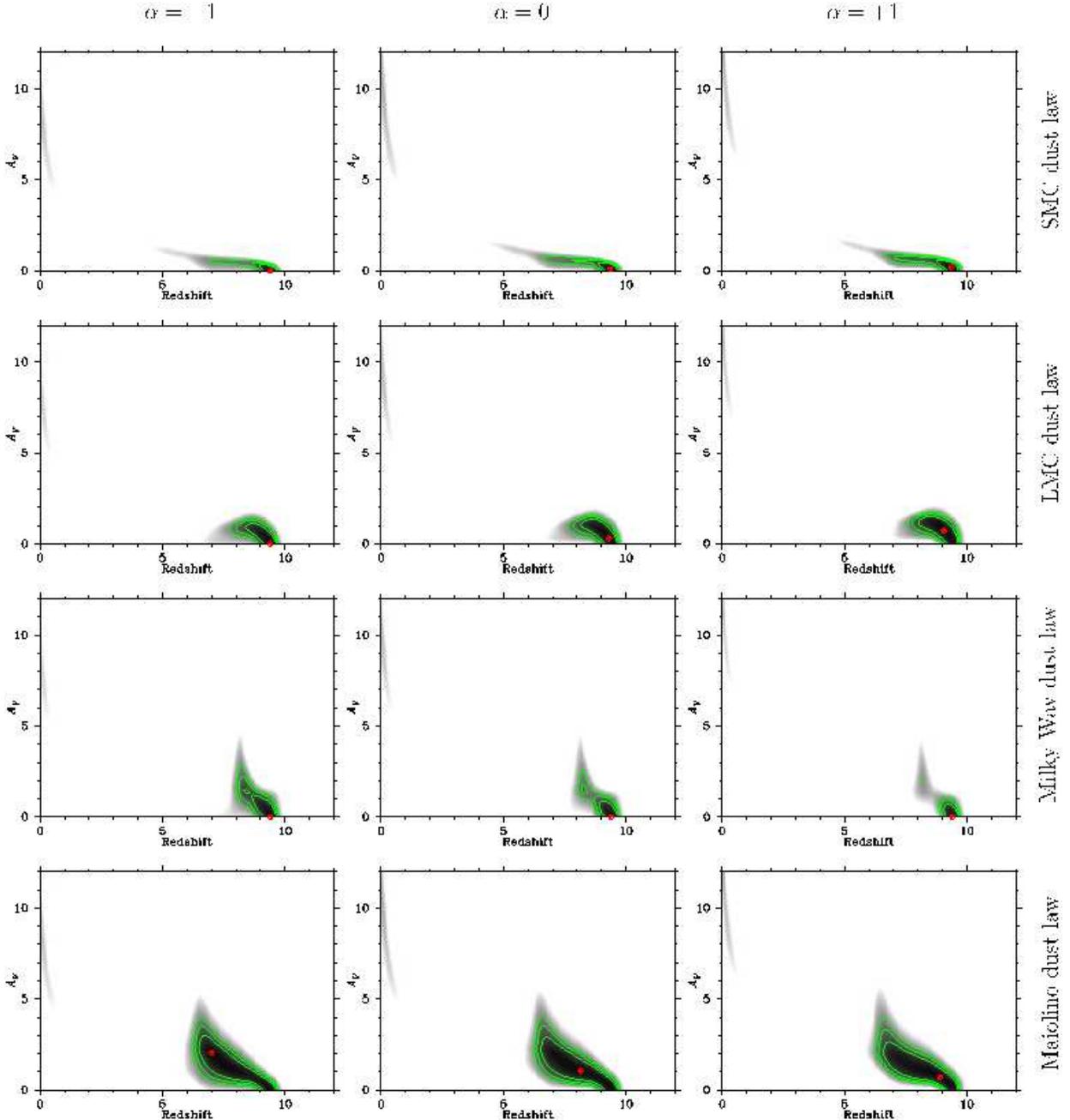}}
\end{center}
\caption{Results of SED fits with a range of possible values for the
temporal power-law index, and different reddening laws. The plots encompass the 
canonical reddening laws for the SMC, LMC and Milky Way (which are 
characterized by the increasing influence of the 2175 \AA\ bump) as well as the
law of \citet{maiolino04} which is approximately flat (``gray") from $\sim1800$ to $3000$ \AA. 
As can be seen, the assumed temporal 
index has only a minimal impact on our results, and our assumption 
of $\alpha = 0$ therefore does not affect our analysis. The majority of GRB afterglows 
are best fit with SMC-like absorption, and we therefore adopt this
as our choice model \citep[e.g.][]{schady07}. Other laws can produce 
broader allowed redshift ranges, in particular 
extending as low as $z \sim 6.3$ at 99\% confidence in the \citet{maiolino04} case, but all rule out low-$z$ ($<6$) scenarios. }
\label{contpanel} 
\end{figure*}

In Figure~\ref{contpanel} we show similar likelihood contours for fits spanning a broader range of models
with different prior assumptions for the temporal power-law decline index $\alpha$, and commonly used
dust laws.  Changing $\alpha$ to $\pm1$ makes rather little difference, and in any case, as
discussed above, there is evidence to suggest the luminosity was not changing
even as rapidly as this.
Varying the dust law does have more effect, largely due to the 2175 \AA\ feature in the MW,
LMC  \citep{pei92}, and \citet{maiolino04} laws producing the blue $H-K$ color even at slightly
lower redshifts, although in most cases the best fit remains $z\gtrsim9$.
The \citet{maiolino04} dust law was determined from observations of a quasar at $z=6.2$
and is argued to consistent with dust produced largely from early supernovae
(note that this law is only defined up to $\sim3200$ \AA\ in the rest frame, and we
therefore graft it to the SMC law at this point).
This  case is interesting as it does allow redshifts as low as 
$z\sim6.5$ at 99\% confidence, although to date,
only GRB\,071025, with a photo-$z\sim5$, has shown  evidence
of requiring such a dust law \citep{Perley:2010cr}.

\begin{figure*}[ht]
\begin{center}
\hspace{-0.2cm}
\resizebox{14.5truecm}{!}{\includegraphics[angle=270]{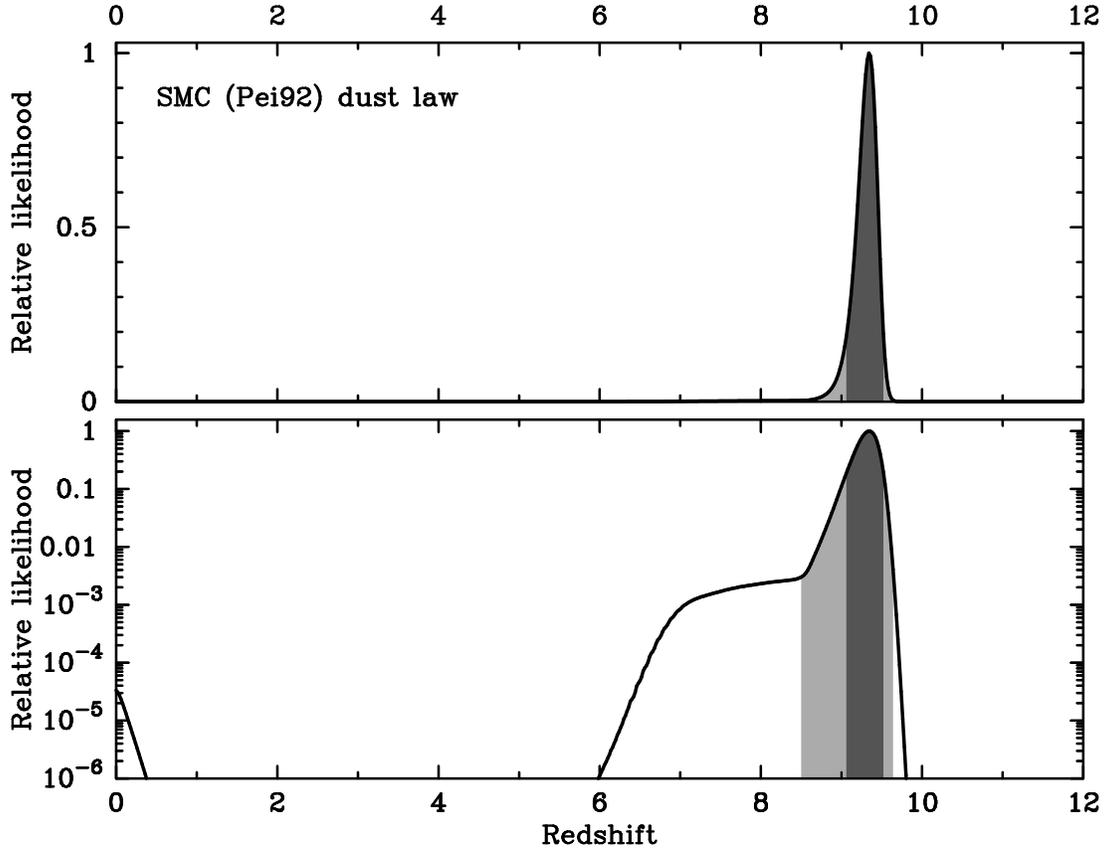}}
\end{center}
\caption{Posterior likelihood plotted on both a linear (upper) and
log (lower) scale, for the models assuming an SMC dust law,
where we have marginalized over both $\alpha$ (assumed to have a flat prior between $-1$ and $+1$) and $A_V$ 
(assumed to have a flat prior between 0 and 12).  The dark and light shaded bands show the extent of
the 90\% and 99\% enclosed likelihood regions respectively.}
\label{1Dprob} 
\end{figure*}

Finally, in Figure~\ref{1Dprob} we show the likelihood as a function of redshift
for the SMC dust-law models
having marginalized over both $\alpha$ (assumed a flat prior between -1 and +1)
and $A_V$ (assumed a flat prior between 0 and 12).  The maximum likelihood 
is at $z=9.38$, and 90\% of the  likelihood is between $9.06<z<9.52$.

\subsection{Implications of the Absence of a Host Galaxy}

\begin{figure*}[ht]
\begin{center}
\resizebox{14truecm}{!}{\includegraphics{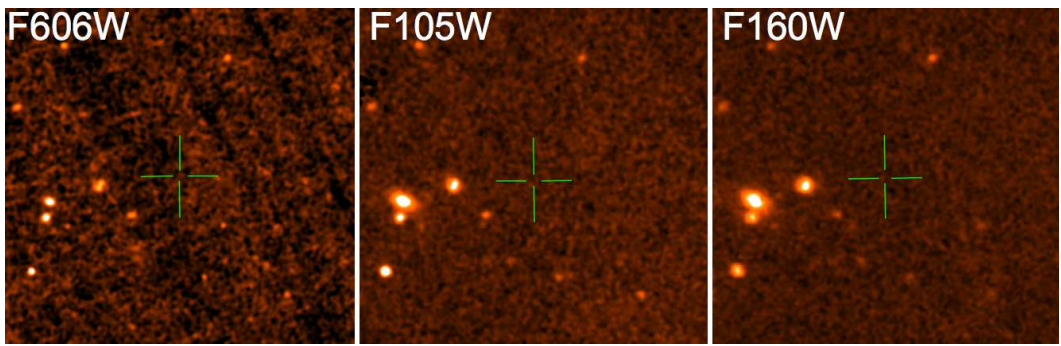}}
\end{center}
\caption{Our late time {\em HST} observations of the GRB\,090429B field in the optical and NIR. 
No host galaxy is detected in any filter, supporting a high-redshift origin, 
since a host with $z<1$ would be very unlikely to be fainter than these
limits, even if dusty.
At $F160W$ the host remains undetected, but the 
observations reach limits which would uncover $\sim 50$\% of the $z>8$ candidates
in the Hubble Ultra-Deep Field (UDF). Hence, the non-detection 
of any host is fully consistent with our high-$z$ model, 
but inconsistent with any lower redshift, high extinction scenario. }
\label{hst} 
\end{figure*}

\begin{figure*}[ht]
\begin{center}
\resizebox{14truecm}{!}{\includegraphics{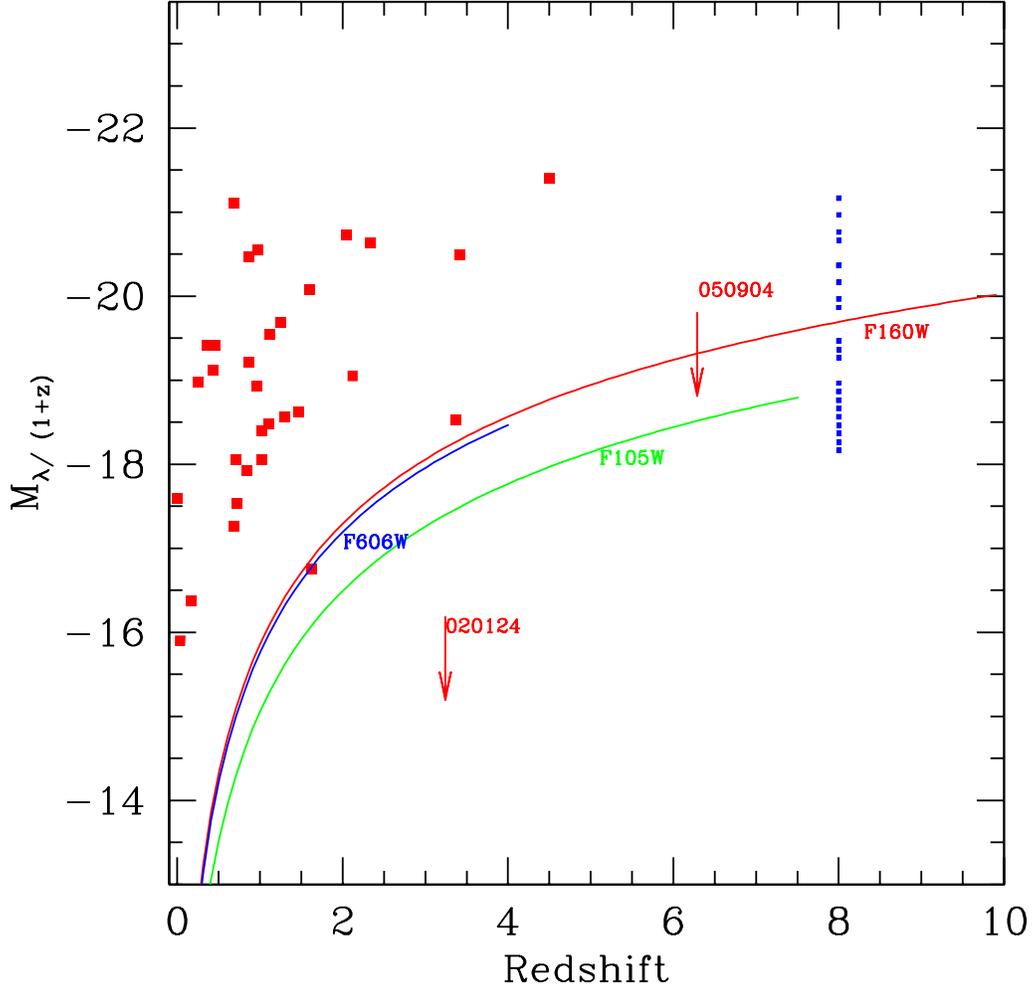}}
\end{center}
\caption{Solid lines show the 3$\sigma$ absolute magnitude limits for GRB 090429B in each of our filters
$F606W$ (blue), $F105W$ (green) and $F160W$ (red). 
The inferred absolute magnitudes (AB) of a sample of 
GRB host galaxies \citep{fruchter06}, as a function of redshift  \citep[plot modified from ][]{perley09a}. The known GRB hosts
with $HST$ observations are plotted as red points, and are supplemented 
at high redshift by the observations of GRB 050904 by \citet{berger06d}. As can be seen, all of these lines lie significantly below the majority of GRB hosts and offer support for
a high-redshift origin for GRB 090429B. The blue points at $z \sim 8$ represents the Lyman break sample of \citet{Bouwens:2010ly}.
As can be seen, the
limiting magnitude for GRB~090429B lies roughly at the median of this distribution, and so the non-detection in our observations would not be unexpected
at \zbest }
\label{mb} 
\end{figure*}

Our late time date taken with Gemini and {\em HST} 
are potentially extremely valuable, since we can use the absence of any host galaxy candidates to assess the plausibility of
any lower-$z$ solutions to our photometric redshifts (the {\em HST} images are shown in Figure~\ref{hst}). 
The detection of a host galaxy in the optical was used, for example, to show that GRB\,060923A 
was $z<3$ despite its afterglow being a $K$-band drop-out \citep{Tanvir:2008wd}.
In the case of GRB\,090429B, the possible low-redshift scenarios seem to be those with $z<1$ and 
high dust extinction $A_V\sim10$
(although we emphasize that such models remain formally ruled out). 
The limits these data provide on this are shown graphically in 
Figure~\ref{mb}, where we plot the absolute inferred magnitude of the host galaxy in the observed $V$, $Y$ and $H$ bands as a function of redshift. For completeness we cut each
line at the point where $1216$\AA $\times (1+z)$ passes the central wavelength of the band. At $z=0.1$ close to the minimum of our
lower redshift solution we obtain inferred absolute magnitude limits of $M_V > -10.6$, $M_{Y} > -9.9$, $M_H > -10.5$, these exceptionally deep limits are comparable
to the luminosities of bright globular clusters, and significantly fainter than any known GRB or supernova host galaxy, indeed
they place limits of $\leq10^{-4}$ $L^*$ \citep{blanton03}. 
Even at $z=1$ the observed $Y$-band limits would imply $M_B > - 15.1$, or $\leq 0.001 L^*$ 
\citep{ryan07}.

In this regard it is worth noting that the lower redshift solutions are only viable in cases where the host galaxy extinction is high, whereas such
faint galaxies typically have low metallicity, and little dust, and it is therefore extremely unlikely that one could create the extinction 
($A_V \sim 10$) necessary to explain GRB 090429B. 
Any $z < 3$ solution would require the host of GRB 090429B to be fainter than the 
large majority of GRB hosts currently known. Furthermore,
our wide wavelength coverage would also allow us to uncover any 
very red dusty host galaxies, which would provide the necessary extinction, but
would be missing from optical only searches 
\citep[e.g.,][]{Svensson:2010fk,Levan:2006}

This offers strong support for our high-$z$ model, In these cases only the {\em F160W} observation yields potential information
as to the magnitude of the host galaxy. The inferred 1500 \AA\ absolute AB magnitude at \zbest is $-19.95$. This lies
roughly in the middle of the observed absolute magnitude distribution of $z \sim 8$ candidate galaxies found in deep {\em HST}
ACS and WFC3 imaging \citep{Bouwens:2010ly}, and  thus the non-detection of a galaxy at \zbest is not
unexpected in observations of this depth.

\subsection{Other Indicators of High Redshift}
In addition to the above discussion there are additional lines of evidence which offer support for the 
high-$z$ interpretation of GRB~090429B. 
The best  $z\lesssim5$  solution  is actually at $z\sim0$
(although it remains a very poor fit to the  available optical data) 
and requires $A_V \sim 10$, for an SMC extinction law,
which would correspond to a foreground $N_{\rm H} \sim 10^{23}$ cm$^{-2}$, nearly two orders of magnitude
larger than is observed. 
While the dust-to-gas ratios observed through the MW can show
moderately large variations, such a large offset would be unheard of, particularly 
in GRB afterglows where typically the ratio of dust extinction to X-ray determined gas column
is actually less than is seen locally \citep{schady10}.  Hence the observed X-ray
spectrum seems to rule out any low redshift ($z\lesssim1$), high extinction scenario. 
This is illustrated in Figure~\ref{cont} which shows that the best low-redshift  solutions 
are well above the contours 
of $A_V$ inferred from the excess $N_{\rm H}$ assuming
typical GRB dust-to-gas ratios.

A second line of evidence comes from observed high-energy correlations seen in many GRBs. In particular
the relation between the peak energy of the $\nu F_{\nu}$ spectrum ($E_p$) and the burst isotropic
energy \citep{Amati:2008fy}.  Although this relation has a significant scatter, it can also be used to place some constraints on the burst
redshift under the assumption that all long bursts should follow the relation. 
For GRB~090429B, the burst is only consistent with this
relation at better than $3\sigma$ if $z>1$, implying that it also disfavors very low redshift models for the origin of
GRB~090429B.  This result couples with the lag-luminosity relation, which would imply an
isotropic luminosity of $L_{\rm iso}\sim 10^{53}$ erg s$^{-1}$, similar to other long GRBs in the
``silver sample''  \citep{Ukwatta:2010lr}.

\subsection{Rest-frame Properties}
\label{sec:rest}
The observed fluence of GRB~090429B is \fluence,  comparable to that observed for GRB~090423, and 
implies an isotropic energy release in the $15-150$~keV band of $E_{\rm iso} = 3.5 \times 10^{52}$\,erg at \zbest. Its
absolute X-ray (rest-frame $\sim$3--100\,keV)
brightness at $\sim$ 1000\,s of $\sim 2 \times 10^{49}$\,erg\,s$^{-1}$, 
and $K$-band luminosity of $M_{\lambda / (1+z)}$ of $\sim -26.2$(AB) are also similar both to GRB~090423 and
the bulk of the long GRB population.

The rest-frame duration at first sight seems surprisingly short, with $T_{90}/(1+z)\sim0.5$\,s.  
Interestingly, three of the four highest redshift GRBs discovered prior to this one
have also had rather low values of $T_{90}/(1+z)$ around 1--4\,s \citep{Ruiz-Velasco:2007kl,Greiner:2009yq,tfl09,sdc09}.
A possible explanation for this tendency may be that for high redshift  sources the BAT is observing
at rest frame MeV energies, where the light curves tend to be more rapidly variable 
and shorter duration than at lower energies,
and thus it is plausible that  just a single peak of emission is detected rising up above the noise in
these cases.

An analysis 
obtained
using the technique discussed in \cite{Ukwatta:2010lr} reveals that
this GRB presents a small value of  
${\rm lag}31/(1+z)=58\pm27$\,ms
between the low-energy (15--25\,keV) and the
high-energy (50--100\,keV) channels.  
The rest-frame correlations found by \citet{Ukwatta:2010lr}
indicate this corresponds to  a peak isotropic
luminosity $L_{\rm iso} >10^{52} \, {\rm erg s^{-1}}$, consistent with the observed
fluence and short duration.

Another interesting issue is that of the absorption inferred from the X-ray spectrum.
Although the measurement is not highly significant, if taken at face value, the rest-frame column 
density is $N_{\rm H}=1.4^{+1.0}_{-1.0}\times10^{23}$\,cm$^{-2}$ (90\% confidence range).
This would be already high compared to most other {\em Swift} observed GRBs,
and would be higher still if, as is very likely, the metallicity is substantially less than
Solar (which by convention is often assumed in calculating $N_{\rm H}$).
Such a high-column density is no doubt surprising, although a similar
value was found for GRB\,090423 \citep{tfl09,sdc09}.  As in that case, it also raises the question
of whether a high gas column would be compatible with the low extinction indicated by the NIR afterglow.
GRBs are expected to be able to destroy dust to fairly large distances from
their birth sites \citep[e.g.,][]{wd00,fkr01}, but the good fit for
many afterglow SEDs with a SMC extinction law suggests they have
not generally been highly modified by dust destruction, which would tend
to produce ``gray" extinction laws \citep[e.g.][]{schady10}.  In any event,
the large error bar on this measurement
and potential systematic uncertainties in calibrating the soft response of the XRT, 
taken together with the wide range of dust-to-gas ratios seen to other GRB sight-lines, makes the significance of this
conflict hard to assess at the present time.


On balance, it seems that in most respects the general properties of
GRB~090429B do not stand far apart from the population of long-duration GRBs, even at the inferred redshift
of \zbest. In particular it shows no evidence that its progenitor is distinct from those of GRBs seen in the more local universe. 
This is of particular importance at \zbest, since it is close to the redshift where WMAP observations imply the
bulk of reionization of the universe occurred  \citep[$z=10.6 \pm 1.2$,][]{Komatsu:2011tg}. 
This reionization process is likely to have been driven by the first generations 
of star formation, including Population III whose pristine H+He composition 
is expected to lead to 
generally more massive stars. 
It has been proposed that a consequence of this
could be that Population III stars produce particularly long duration, and energetic GRBs \citep[e.g.][]{Meszaros:2010dp}. 
This is clearly not the case for
GRB~090429B, and 
hence we conclude that its progenitor was more likely to be a high-mass second generation 
(Population II) star. 


\section{Conclusions}
\label{sec:sum}

We have presented our discovery and multi-wavelength observations of
GRB\,090429B and its NIR afterglow, and a deep late-time search for its host galaxy. 
The afterglow exhibited a strong spectral break  in the $J$ band, which coupled with
the non-detection in the optical, and relatively blue $H-K$ color, allows
us to derive a best-fit photometric redshift of \zbest. 
It is, of course, important to look carefully at the evidence
against a lower redshift origin, since we know that {\em Swift} GRBs
exist in much greater number at $z<4$ than above it.
Our afterglow photometry allows us to exclude all $z<6$ solutions with high confidence.
A low redshift ($z\lesssim1$) is also effectively ruled out by our {\em HST} observations which would
easily locate such dusty galaxies in either the optical or IR at $z<1$, and also the
relatively modest excess $N_{\rm H}$ which would not be consistent with a very high dust
column. 
This immediately implies that GRB~090429B is one
of the most distant objects yet discovered. 
The maximum-likelihood solution, with our preferred assumption of 
an SMC extinction law, is $z=9.38$ with a
 90\% likelihood range of $9.06<z<9.52$. The conclusions do not depend
 sensitively on the priors adopted for other parameters, although a \citet{maiolino04} dust
 law would favor somewhat lower redshift, but still $z>7$, at the expense of requiring 
 fairly significant extinction (up to rest frame $A_V\sim2$).
Since all $z>6$ bursts
observed to date are consistent with having $A_V = 0$ \citep{Zafar2010,Zafar2011}, 
this suggests that \zbest provides a good estimate of the redshift of GRB\,090429B. 

Our campaign shows again how rapid-response multiband NIR observations
play a crucial role in identifying candidate extreme-redshift
afterglows. However, it also highlights the need for even more rapid
observations and decisions to maximize the likelihood that spectroscopic observations
can be successfully obtained. In the
future, additional dedicated ground-based optical/NIR multi-band
imagers such as GROND and RATIR  \citep{farah:77357Z} can be expected to feed
further such candidates directly to 
NIR spectrographs including X-Shooter on the VLT \citep{DOdorico:2004uq}, FIRE on
Magellan \citep{Simcoe:2008qy}, and GNIRS on Gemini \citep{Elias:2006fk}; ultimately,
such prompt spectroscopy of extreme-redshift candidates will not only
resolve the nature of these events, but quite likely succeed in
realizing the extraordinary promise of GRBs as probes of the
extreme-redshift universe.

\acknowledgements

The Gemini data, acquired under the program ID GN-2009A-Q-26, are
based on observations obtained at the Gemini Observatory, which is
operated by the Association of Universities for Research in Astronomy,
Inc., under a cooperative agreement with the NSF on behalf of the
Gemini partnership:  the National Science Foundation (United 
States), the Science and Technology Facilities Council (United Kingdom), the 
National Research Council (Canada), CONICYT (Chile), the Australian Research 
Council (Australia), Minist\'{e}rio da Ci\^{e}ncia e Tecnologia (Brazil) 
and Ministerio de Ciencia, Tecnolog\'{i}a e Innovaci\'{o}n Productiva (Argentina).
Based on observations made with the NASA/ESA $Hubble Space Telescope$, 
obtained from the data archive at the Space Telescope Institute. STScI is operated by 
the association of Universities for Research in Astronomy, Inc. under the NASA contract  NAS 5-26555.
Data presented in this paper is associated with programme  GO-11189.
Part of the funding for
GROND (both hardware as well as personnel) was generously granted from
the Leibniz-Prize to Prof. G. Hasinger (DFG grant HA 1850/28-1).  TK
acknowledges support by the DFG cluster of excellence ``Origin and
Structure of the Universe''. AR acknowledges funding from the Science and Technology Funding Council.
The Dark Cosmology Centre is funded by the Danish National Research Foundation.
FOE acknowledges funding of his Ph.D.
through the \emph{Deutscher Akademischer Austausch-Dienst} (DAAD).
KT and XFW acknowledge NASA NNX09AT72G and NNX08AL40G.

We thank Paul Hewett for helpful discussions about absolute infrared flux calibration.


\bibliographystyle{apj_8}
\bibliography{journals_apj,grboldbi,bibi090429B,bibthesis}

%
%

\begin{deluxetable}{lll} 
\tablecolumns{3} 
\tablewidth{0pc}
\tablecaption{X-ray Observations} 
\tablehead{ 
\colhead{$T-T_0$ (s)}  &\colhead{Flux Density ($\mu$Jy)}   & \colhead{Error ($\mu$Jy)}    }
\startdata 
158 & 0.53 & 0.12 \\
 291 & 0.80 & 0.18\\
 368 &1.18 & 0.27\\
 452 &1.19 & 0.27\\
 561 & 1.69 & 0.36\\
 651 & 1.37 & 0.31\\
 708 & 1.35  & 0.30\\ 
768 & 1.18  & 0.27\\
838 & 1.00 & 0.22\\
927 & 1.31  & 0.21\\
4618 & 0.099 & 0.026\\ 
5636 & 0.111&  0.029\\
6417 & 0.126 & 0.033\\
10316 & 0.089  & 0.023\\
10982 & 0.087 & 0.023\\
15921 & 0.0269  & 0.0053\\
120249 & 0.0021 & 0.0006\\
\enddata 
\tablecomments{X-ray observations obtained by the XRT instrument
  onboard the \swift\ satellite.  The flux density is calculated at 2\,keV,  while 
  the conversion factor from flux to 
  Jy is $8.87\times 10^{4}$ erg cm$^{-2}$ s$^{-1}$ Jy$^{-1}$ (0.3--10\,keV). 
  The counts to flux conversion factor is $2.984\times10^{-11}$ erg cm$^{-2}$ s$^{-1}$ count$^{-1}$.
  Uncertainties are 1$\sigma$.
  For the best fit X-ray spectrum 
  parameters see Section \ref{sec:obs}.}
\label{tab:2table2}
\end{deluxetable}

\begin{deluxetable}{ccccc} 
\tablecolumns{5} 
\tablewidth{0pc}
\tablecaption{Log of Ground-based Optical/NIR Observations} 
\tablehead{ 
\colhead{$T-T_0$ (s)}  &\colhead{Magnitude}   & \colhead{Flux Density ($\mu$Jy)} & \colhead{Filter} & \colhead{Telescope} }
\startdata 
990	& $> 23.08 $ & &\gp & GROND\\
990	& $> 22.86 $ & &\rp & GROND\\
990	& $> 22.03 $ & &\ip & GROND \\
990	& $> 21.87 $ & &\zp & GROND\\
990	& $> 21.06 $ & &$J$ & GROND \\
990	& $> 20.50 $ & &$H$ & GROND \\
990	& $> 19.84 $ & &$K$ & GROND \\
3224 & $>24.5$ & $0.00 \pm 0.20$ & $B$ & VLT/FORS2 \\
4017 & $> 25.9 $ & $-0.08 \pm 0.08$  & $R$ & VLT/FORS2 \\
5144 & $>23.6  $ & $0.27 \pm0.34$  & $z$ & VLT/FORS2 \\
8135	& $>25.7 $ & $0.02\pm0.06$ & \ip & Gemini-N/GMOS\\
9350	& $>24.5  $ & $0.02\pm0.18$ &\zp & Gemini-N/GMOS \\
10611	& $22.80 \pm 0.16$ & $2.82\pm0.44$ & $J$ & Gemini-N/NIRI \\
11785	& $21.41 \pm 0.05 $ & $10.21\pm0.50$ &$H$ & Gemini-N/NIRI\\
13280	& $21.12\pm 0.04 $ & $13.26\pm0.51$ &$K$  & Gemini-N/NIRI\\
95658      & $22.42 \pm 0.16   $  & $4.0\pm0.6$ &  $K$ & Gemini-N/NIRI \\
$1.2 \times 10^{6}$  &$> 27.07$ & & \rp & Gemini-N/GMOS\\ 
\enddata 
\tablecomments{Optical/NIR observations of \grb. Magnitudes
  are quoted in the AB system, and corrected for the expected Galactic
  extinction along the line of sight, $E_{B-V} = 0.015$.  Quoted errors are 1$\sigma$
  and limits are 
  at the 3$\sigma$ level.}
\label{tab:2table1}
\end{deluxetable}

\begin{table}[htdp]
\caption{Log of {\em HST} Observations of the GRB 090429B Field}
\begin{center}
\begin{tabular}{llllll}
\hline
Date & Start Time & Inst/Filter & Exp time & Limit & Flux density ($\mu$Jy) \\
\hline
3 Jan 2010 & 03:13 & ACS/F606W & 2100 & $>27.6$ & $0.005 \pm 0.008$\\
10 Jan 2010 & 21:54 & WFC3/F160W & 2412 &  \\
22 Feb 2010 & 19:22 & WFC3/F160W & 2412 & $>27.5$ & $0.007 \pm 0.005$  \\
24 Feb 2010 & 03:19 & WFC3/F105W & 2412 & \\
28 Feb 2010 & 13:56 & WFC3/F105W & 2412  & $>28.3$ & $-0.001 \pm 0.005$\\
\hline
\end{tabular}
\end{center}
\tablecomments{A log of the $HST$ optical and NIR observations of the GRB~090429B field. 
Flux densities are given in the measured apertures and are not corrected for light outside the
apertures.  Errors are 1$\sigma$ and the
limits are given in the AB-magnitude system at the $3\sigma$ level (and
do include aperture corrections). 
}
\label{hstdata}
\end{table}%

\begin{deluxetable}{lllllllll} 
\tablecolumns{8} 
\tablewidth{0pc}
\tablecaption{Secondary Standards within the NIRI Field of View} 
\tablehead{ 
\colhead{Star} &  \colhead{RA} & \colhead{DEC} & \colhead{J$_{2MASS}$} & \colhead{J$_{cal}$} & \colhead{H$_{2MASS}$} & \colhead{H$_{cal}$} & \colhead{K$_{2MASS}$} & \colhead{K$_{cal}$} } 
\startdata
A &  14:02:35.05 &   32:11:07.2 &    14.754  &   $14.753 $ &   $14.411$ &   $14.407$ &  $14.212$  &   $14.341$ \\
   &	                       &                       & $\pm 0.034$ & $\pm 0.006$ & $ \pm 0.055$ & $ \pm 0.008$ & $\pm 0.075$ & $ \pm 0.010$ \\
  B &   14:02:40.60 &  32:09:28.9 &   $15.419$ &   $15.451$ &   $14.968$ &  $15.023$ &   $14.846$ &   $14.944 $ \\
  & 			&			& $ \pm 0.055$ & $ \pm 0.007$ & $\pm 0.076$ & $ \pm 0.009$ & $\pm 0.127$ & $\pm 0.015$ \\
  C &   14:02:38.11 &  32:10:08.9 &                   &   $19.523$ &                       &  $19.001$ &            &           $ 18.585$ \\
  & 			&			&  & $ \pm0.035$ &  & $ \pm0.028$ & & $\pm0.032$ \\
\enddata
\tablecomments{Vega magnitudes for our two secondary standard stars utilized 
in photometry of the afterglow of GRB 090429B. The 2MASS entries refer to the
magnitudes contained within the 2MASS catalog, while those denoted ${cal}$ refer 
to our improved values based on the UKIRT/WFCAM and ESO2.2/GROND observations. Uncertainties are 1$\sigma$.}
\label{crosscalib}
\end{deluxetable}


%
%

\end{document}